\documentstyle[preprint,revtex]{aps}
{\makeatletter%
\gdef\labeleqs#1{{%
\edef\@currentlabel{%
\ifappendixon\appletter\fi
\ifsecnumbers\ifnum\c@secnum>0
\arabic{secnum}.\fi\fi\arabic{equation}}%
\label{#1}%
}}%
}
\begin{document}
\draft
\preprint{IFUP-TH 35/92}
\begin{title}
Monte Carlo simulation of lattice ${\rm CP}^{N-1}$ models
at large $N$.
\end{title}
\author{Ettore Vicari}
\begin{instit}
Dipartimento di Fisica dell'Universit\`a and
Istituto Nazionale di Fisica Nucleare, I-56126 Pisa, Italy
\end{instit}
\begin{abstract}

In order to check the validity and the range of applicability
of the $1/N$ expansion, we performed numerical simulations of the
two-dimensional lattice ${\rm CP}^{N-1}$ models at large $N$,
in particular we considered the ${\rm CP}^{20}$ and the
${\rm CP}^{40}$ models.

Quantitative agreement with the large-$N$ predictions
is found for the correlation length
defined by the second moment of the correlation function,
the topological susceptibility and the string tension.
On the other hand, quantities involving the mass gap
are still far from the large-$N$ results showing a very slow approach
to the asymptotic regime.

To overcome the problems coming from the
severe form of critical slowing down observed at large $N$
in the measurement of the
topological susceptibility by using standard local algorithms,
we performed our simulations implementing the Simulated Tempering method.

\end{abstract}

% ========================= BODY =========================

\narrowtext
\section{Introduction}
\label{Introduction}

The most attractive feature of two dimensional $CP^{N-1}$ models
is their similarity with the Yang-Mills theories in four space-time
dimensions.
Most properties of ${\rm CP}^{N-1}$ models
have been obtained in the context of the $1/N$ expansion
around the large-$N$ saddle point solution
\cite{DiVecchia-Luscher,Witten,CPN}.

An alternative and more general non-perturbative approach
is the simulation of the theory on the lattice.
Recently there has been considerable interest in simulations of
lattice ${\rm CP}^{N-1}$ models
\cite{CPNlatt,CPNlatt_2,Wiese,Hasenbusch,Hasenbusch2,Michael,%
Michael2,Wolffcp3,Muller}.
Simulations up to $N=10$ have shown a qualitative agreement
with some of the features
derived from the continuum $1/N$ expansion, especially
those concerning the sector of theory closely connected
to the dynamically generated gauge fields, such
as topology and confinement.
On the other hand,
while the $1/N$ expansion predicts a complex mass spectrum,
evidence of other bound states, beyond the fundamental
one in the adjoint positive parity channel, is not found up to
$N=10$ \cite{CPNlatt_2}.
This is not a surprise,
in that the agreement with the $1/N$ expansion can only
be reached at very large $N$, because of the very large coefficient
in the effective expansion parameter $6\pi/N$ that can be
extracted from a nonrelativistic Schr\"odinger equation analysis of
the linear confining potential \cite{Witten,CPN}.
It is then important to obtain numerical results
at large $N$ to check the validity and the range of applicability of the
$1/N$ expansion.

In this paper we present the results of some simulations of
the lattice ${\rm CP}^{N-1}$ models at large $N$, in  particular
$N=21$ and $N=41$, and we compare them to the large-$N$ predictions
coming from the $1/N$ expansion.

At large $N$ a particularly severe form of
critical slowing down has been
observed in measuring the topological susceptibility $\chi_t$
by using local updatings.
For $N=10$ the autocorrelation time of $\chi_t$ seems to grow exponentially
with respect to the correlation length \cite{CPNlatt}.
Moreover, this phenomenon becomes stronger with increasing $N$,
making the simulations effectively non ergodic, already at small $\xi$.
In order to overcome this difficulty
and to perform simulations sampling correctly
the topological sectors,
we used the Simulated Tempering method proposed
by Marinari and Parisi \cite{ST}.
In this method the temperature becomes a dynamical variable,
and it is changed while keeping the system at equilibrium.

This paper is organized as follows.

In Sec.\ \ref{Lattice} the lattice actions adopted for
numerical simulations and the lattice definitions
of physical observables are presented.

In Sec.\ \ref{Monte} we describe the Monte Carlo algorithm.

In Sec.\ \ref{simulation} we present the numerical results
comparing them with the large-$N$ predictions.

\section{Lattice formulation}
\label{Lattice}

We regularize the theory on the lattice by considering
the following action:
\begin{equation}
S_{\rm g} = -N\beta\sum_{n,\mu}\left(
   \bar z_{n+\mu}z_n\lambda_{n,\mu} +
   \bar z_nz_{n+\mu}\bar\lambda_{n,\mu} - 2\right),
\label{basic}
\end{equation}
where $z_n$ is an $N$-component complex scalar field, constrained by
the condition
%\begin{mathletters}
%\begin{equation}
$\bar z_nz_n = 1$,
%\end{equation}
and $\lambda_{n,\mu}$ is a ${\rm U}(1)$ gauge field satisfying
%\begin{equation}
$\bar\lambda_{n,\mu}\lambda_{n,\mu} = 1$.
%\end{equation}
%\end{mathletters}
We also considered its tree Symanzik improved counterpart
$S_{\rm g}^{\rm Sym}$ \cite{Symanzik,CPNlatt} to test universality.
Tests of rotation invariance and stability of dimensionless ratios of
physical quantities showed that $S_g$ and $S_g^{\rm Sym}$ lead to scaling for
rather small correlation lengths \cite{CPNlatt}.

An important class of observables can be constructed by considering
the local gauge-invariant composite operator
\begin{equation}
P_{ij}(x) = \bar z_i(x) z_j(x)
\label{P-ij}
\end{equation}
and its group-invariant correlation function
\begin{equation}
G_P(x) = \left<\mathop{\rm Tr} P(x) P(0)\right>_{\rm conn} \,.
\label{G-P}
\end{equation}

The standard correlation length $\xi_{\rm w}$ is extracted from the
long-distance behavior of the zero space momentum correlation
function (``wall-wall'' correlation).
$\xi_{\rm w}$ should reproduce in the continuum limit the inverse mass gap,
that is the inverse
mass of the lowest positive parity state belonging to the adjoint
representation.

An alternative definition of the correlation length $\xi_G$ comes
from considering the second moment of the correlation function $G_P$.
In the small momentum regime we expect the behavior
\begin{equation}
\widetilde G_P(k) \approx
{Z_P \over \xi_G^{-2} + k^2}\,,
\end{equation}
where $\widetilde G_P(k)$ is the Fourier transform of $G_P(x)$.
The zero component of $\widetilde G_P(k)$ is by definition the
magnetic susceptibility $\chi_{\rm m}$.
On the lattice from the two lowest
components of $\widetilde G_P(k)$ we can derive
the following definition of $\xi_G$:
\begin{equation}
\xi_G^2 = {1\over4\sin^2\pi/L} \,
\left[{\widetilde G_P(0,0)\over\widetilde G_P(0,1)} - 1\right].
\label{xiG}
\end{equation}
In the scaling region the ratio $\xi_G/\xi_{\rm w}$ must be a
constant, scale-independent number.

For $N=2$ $\xi_G/\xi_{\rm w}\simeq 1$ within 1\% \cite{CPNlatt},
while the large-$N$ expansion predicts \cite{CPN-lett}
\begin{equation}
{\xi_{\rm G}\over\xi_{\rm w}}=\sqrt{{2\over 3}}\,+\,
O\left( {1\over N^{2/3} } \right)\;.
\label{ratioxi}
\end{equation}

The mass gap, and therefore $\xi_{\rm w}$, is
a non-analytic function of $1/N$ around $N=\infty$
depending on $N^{-2/3}$. Instead  $\xi_{\rm G}$
can be expand in power of $1/N$ \cite{CPN}:
\begin{equation}
(\xi_G\Lambda_{SM})^{-1}\,=\,
\sqrt{6} \left[ 1 + {6.1325\over N} + O\left({1\over N^2}\right)\right]
\label{xiG/Lambda}
\end{equation}
where $\Lambda_{SM}$ is the $\Lambda$ parameter of the
sharp-momentum cut-off regularization scheme \cite{sharp}.
Standard perturbative calculations give
$\Lambda_{SM}/\Lambda_g\,=\,\sqrt{32}\exp (\pi/2N)$,
where $\Lambda_g$ is the $\Lambda$ parameter of the
lattice action $S_g$.

The quantity $Z_P = \chi_{\rm m}\xi_G^{-2}$ is
related to the renormalization of the composite operator
$P_{ij}$. Its dependence on $\beta$ can therefore be determined by
renormalization group considerations.
One finds that
\begin{equation}
Z_P = c \beta^{-2} \left[1 + O\left(1\over N\beta\right)\right]\,,
\label{Z_P}
\end{equation}
where $c$ is a constant independent of the regularization scheme
and therefore of the lattice action.
In the large-$N$ limit it turns out to be \cite{CPNlatt_2}
\begin{equation}
c = {3\over 2\pi}\;\left[1 + {8.5414\over N} +
O\left(1\over N^2\right)\right]\,.
\label{cc}
\end{equation}

Another important class of observables is that connected to
the dynamically generated gauge field, such as the topological
susceptibility and the string tension.

The geometrical definition of the topological charge is
\cite{Berg-Luscher}
\FL
\begin{eqnarray}
q_n = {1\over2\pi}\,\mathop{\rm Im}\bigl\{
     &&\ln[\mathop{\rm Tr} P_{n+\mu+\nu}P_{n+\mu}P_n]   \nonumber \\
  && + \ln[\mathop{\rm Tr} P_{n+\nu}P_{n+\mu+\nu}P_n]\bigr\},
\qquad \mu\ne\nu \,.
\label{geometrical-Q}
\end{eqnarray}
The topological susceptibility should then be extracted by
measuring the following expectation value
\begin{equation}
\chi_{\rm t}  =
{1\over V} \left< \left(\sum_n q_n\right)^2 \right>\;.
\label{geomtop}
\end{equation}
For large $N$ this definition is
expected to reproduce the physical topological susceptibility
\cite{DiVecchia-Rossi,CPNlatt_2}.

The large-$N$ predictions concerning the topological susceptibility
are \cite{CPN-lett}
\begin{equation}
\chi_{\rm t}\xi_G^2 = {1\over 2\pi N}\,\left(1-{0.3801\over N}\right) +
O\left({1\over N^3}\right)\,,
\label{chipred}
\end{equation}
and \cite{Luscher-lett}
\begin{equation}
\chi_{\rm t}\xi_{\rm w}^2 = {3\over 4\pi N}\,+\,
O\left({1\over N^{5/3}}\right)\,.
\label{chipredluscher}
\end{equation}
Eq.\ (\ref{chipred}) and Eq.\ (\ref{chipredluscher}) are not in
contradiction with each other due to Eq.\ (\ref{ratioxi}),
but the first one should be testable at lower values of $N$ according
to the powers of $N$ in the neglected terms.

The large-$N$ expansion predicts an
exponential area law behavior for sufficiently large Wilson
loops \cite{CPN}:
\begin{equation}
W({\cal C}) = \prod_{n,\mu\in{\cal C}} \lambda_{n,\mu}
\sim e^{-\sigma A({\cal C}) - \rho P({\cal C})}
\quad {\rm for} \quad A({\cal C}) \gg \xi^2 \,,
\end{equation}
where $\sigma$ is the Abelian string tension and $\rho$ is a
perimeter term.
This implies also that the dynamical matter fields
do not screen the linear potential at any distance.
Monte Carlo simulations at $N=4$ and $N=10$ confirmed the
absence of screening effects \cite{CPNlatt_2}.
The large-$N$ prediction for $\sigma$ is
\begin{equation}
\sigma\xi_G^2\,=\,{\pi\over N} \,+\,O\left( {1\over N^2}\right)\;.
\label{sigmapred}
\end{equation}

The string tension can be easily extracted by measuring the Creutz
ratios defined by
\begin{equation}
\chi(l,m) = \ln {W(l,m{-}1)\,W(l{-}1,m)
   \over W(l,m)\,W(l{-}1,m{-}1)} \,.
\end{equation}
In a 2-d finite lattice with periodic boundary conditions, the
large abelian Wilson loops of a confining theory are subject to large
finite size effects.
For sufficiently large $R$ the behavior
of the Creutz ratios $\eta(R)\equiv \chi(R,R)$,
i.e. of those with equal arguments,
should be \cite{CPNlatt_2}:
\begin{equation}
\eta(R) \simeq \sigma\,\left[ 1-\left({2R-1\over L}\right)^2\right]\;,
\label{Crratcpn}
\end{equation}
where $L$ is the lattice size.
To compare data from different lattices it is convenient
to define a rescaled Creutz ratio
\begin{equation}
\eta_{\rm r}(R) = \eta(R)\,
    \left[ 1-\left({2R-1\over L}\right)^2\right]^{-1}
\simeq \sigma\;.
\label{Crratres}
\end{equation}

\section{The Monte Carlo algorithm}
\label{Monte}

In most of our simulations we used the simulated tempering method
proposed in Ref.\ \cite{ST}.
The basic idea of this method consists in enlarging the configuration
space of the system by including the temperature, and changing it
while remaining at statistical equilibrium.
Considering a finite set of temperatures $\beta_i$, $i=1,...N_\beta$,
the probability distribution is chosen to be
\begin{equation}
P(\beta_i,x)\,=\,e^{-K(\beta_i,x)}\;,
\end{equation}
where
\begin{equation}
K(\beta_i,x)\,=\,\beta_i H(x) - g_i\;,
\end{equation}
where $x$ indicates the lattice variables, $H(x)$ is the
hamiltonian of the statistical system, and $g_i$ is independent
of $x$.
The probability distribution induced by $K(\beta_i,x)$, restricted
to the subspace $i$, is the Gibbs distribution for $\beta=\beta_i$.

By making the choice
\begin{equation}
g_i\,=\,\beta_i F_i\;,
\end{equation}
where $F_i$ is the free energy at $\beta_i$, the probability
of having a given value of $i$ becomes independent of $i$,
i.e. $P_i=1/N_\beta$.

In practice the simulated tempering method is implemented by performing
the following cycle \cite{ST}:
(i) updating the lattice variables at the temperature $\beta_i$
by using a standard algorithm;
(ii) updating the temperature according to the probability
\begin{equation}
P(\beta_i)\,=\,e^{-\beta_i E + g_i}\;,
\end{equation}
where $E$ is the energy of the configuration obtained in (i).
The expectation values at a given $\beta$ can be obtained performing
the measurements when $\beta_i=\beta$.

In the presence of free energy barriers separating different regions
of the configuration space, the visits of the system
to lower values of $\beta$ will make easier to jump, in that
at lower $\beta$ free energy barriers are lower.

It is important to choose the values of $\beta_i$
so that the probability transition from one value of $\beta$
to another is not negligible. This can be achieved by requiring
a non-negligible overlap in the values of the energy of the configurations
coming from simulations at contiguous values of $\beta_i$,
and using the following approximation for $g_i$:
\begin{equation}
g_{i+1}\,\simeq
\,g_i \,+\,(\beta_{i+1}-\beta_i)\left({E_{i+1}+E_i\over 2}\right)\;,
\label{g_m}
\end{equation}
which is a good approximation when $\Delta \beta$
is small and it is simple to estimate.

In the case of the lattice ${\rm CP}^{N-1}$ models,
going to lower $\beta$ should make it easier to jump from one
topological sector to another, and when the temperature will
decrease again the system will be visiting a different topological
sector with the correct equilibrium probability, providing
us with a well representative ensemble of configurations at the given
value of $\beta$.

To update the lattice variables,
we chose the over-heat bath algorithm \cite{Petronzio-Vicari}
because it is very efficient
in decorrelating the energy and, at the same time,
contains a procedure of overrelaxation.
Furthermore it requires less computational effort
than a standard heat bath.
The implementation of the over-heat bath method in the
lattice ${\rm CP}^{N-1}$ models
is described in Ref.\ \cite{CPNlatt}.

The difficulty in applying the simulated tempering method to the
${\rm CP}^{N-1}$ models is that with
increasing $N$ the fluctuations of the energy
tend to be frozen (at $N=\infty$
only one configuration contributes to the path integral)
making necessary to keep the difference between contiguous $\beta_i$
very small. Therefore in order to work with a wide range of temperatures
we must introduce a large number of $\beta_i$.

We performed also some standard simulations by employing algorithms
consisting in mixtures of over-heat bath and microcanonical
algorithm \cite{CPNlatt}.

\section{Simulations}
\label{simulation}

In Table \ref{ST-table} we present a summary of the runs done
by using the simulating tempering method.
In each run we performed the measurements at two values of $\beta$,
which can be read in Table \ref{datarun-table}.
In Table \ref{datarun-table} we also give a summary of the runs done by
using standard algorithms.
Some preliminary results of the lattice ${\rm CP}^{20}$ model were
already presented in Ref.\ \cite{CPNlatt}. There
the exponential growth of the autocorrelation time of $\chi_t$
allowed to obtain
meaningful measurements of $\chi_t$ only at
small correlation length, $\xi_G\simeq 2.5$,
while simulations at larger $\xi$ did not sample correctly
the topological sectors.
By using the simulated tempering method we performed
simulations up to $\xi_{\rm G}\simeq 4.2 $ obtaining
reliable measurements of $\chi_t$. For the ${\rm CP}^{40}$
model we performed a simulated tempering run with correlation
lengths up to $\xi_{\rm G}\simeq 2.5$.
All simulations were performed setting periodic boundary conditions.

Since the measurements required much more computational time than
the updating procedure and we were essentially interested
in decorrelating the topological charge,
when using the simulated
tempering method  we checked
the value of $\beta$ every 4--5 sweeps  and
performed  the measures when $\beta_i=\beta$.
Errors were estimated by a blocking procedure.
Measurements at different values of $\beta$ but from the same simulated
tempering simulation are not completely decorrelated,
especially those regarding the topological susceptibility.
In the standard runs the integrated autocorrelation times of the
magnetic susceptibility were small,
instead those relative to the topological susceptibility
were very large.
For $N=21$ and at $\beta=0.65$, we found $\tau^{\chi_m}_{int}=3.2(1)$
and $\tau^{\chi_t}_{int}\simeq 100$; by using $S_g^{\rm Sym}$ and at
$\beta=0.60$ $\tau^{\chi_t}_{int}\simeq 600$.
%%$\tau_{\chi_m} \mathop<\limits_\sim 7$.

For large enough $N$, the finite size effects should be dominated
by the size of the ground state and not by its mass.
The $1/N$ expansion predicts a radius of the ground state proportional
to $\xi N^{1/3}$.
A comparison of the finite size scaling functions of $\chi_m$
and $\xi_G$ at $N=10$
and $N=21$ has shown that $z=L/\xi_G\simeq 4.5\,N^{1/3}$ should be a
safe value in order to have finite size effects smaller
than 1\% (at least for $\chi_m$ and $\xi_G$) \cite{CPNlatt}.
We checked this further for $N=41$ by comparing the results
obtained at $\beta=0.57$ on lattices with $L=33$
($z\simeq 16.5$) and with $L=42$ ($z\simeq 21$), and
finding agreement within errors of about 0.5\%.

In Table \ref{some-table} we list the correlation length
$\xi_G$, the ratio $\xi_G/\xi_{\rm w}$,
the dimensionless quantity $\chi_t\xi_G^2$ and
the combination $\beta^2 Z_P\equiv \beta^2 \chi_m \xi_G^{-2}$.
All these quantities were analyzed using the jackknife method.

$\xi_{\rm w}$ was obtained by fitting the wall-wall correlations
starting from a minimum distance $x_{min}$.
We set $x_{min}\simeq 2\xi_{\rm w}$
for the ${\rm CP}^{20}$ model and $x_{min}\simeq 3\xi_{\rm w}$
for the ${\rm CP}^{40}$ model; fits using larger $x_{min}$
gave consistent results.

At all values of $\beta$ we performed a test of rotation invariance by
comparing $\xi_{\rm w}$ with
the correlation length $\xi_{\rm d}$ extracted
from the long-distance behavior of the
diagonal wall-wall correlations of $P_{i,j}$ \cite{CPNlatt}.
We found $\xi_{\rm d}/\xi_{\rm w}\simeq 1$ within errors of about
0.5\% in all cases.

In Fig. \ref{ratio_xi-plot} the ratio $\xi_G/\xi_{\rm w}$
is plotted versus $\xi_G$. We note that at $N=41$
$\xi_G/\xi_{\rm w}$ is still far from the large-$N$ prediction
(\ref{ratioxi}),
indicating a very slow approach to the large-$N$ asymptotic regime.

Data for $\beta^2 Z_P$ show scaling and, for the ${\rm CP}^{20}$ model,
the two actions give close values.
The small discrepancies can
be imputed to the non-universal terms of order $(N\beta)^{-1}$ in Eq.\
(\ref{Z_P}).  The comparison with Eq.\ (\ref{cc}),
which gives $c=0.6717$ for $N=21$ and $c=0.5769$ for $N=41$, is
satisfactory.

At large $N$ the dynamically generated gauge field contains essentially
two distinct types of modes at large distance:
the gaussian fluctuations around the large $N$ saddle point
solution, which are  responsible for confinement, and those
determining the topological properties.
We expect to find in the ${\rm CP}^{N-1}$ models a phenomenon
similar to that observed in the 2-d $U(1)$ gauge model, that is
a large decoupling between the gaussian modes and the topological
ones \cite{multigrid}.
This picture is supported by
the agreement found in the results corcerning observables
not related to the topological properties,
obtained by the simulated tempering
method and by standard simulations which did not sample
correctly the topological sectors, whose results were
reported in Ref. \cite{CPNlatt}.

In Fig. \ref{chit-plot} we plot the dimensionless quantity
$\chi_t\xi_G^2$.
For both the ${\rm CP}^{20}$ and ${\rm CP}^{40}$ models
data are consistent with the large-$N$ prediction
(\ref{chipred}).
In order to quote a value for $\chi_t$, we fitted to a constant
the data selected by taking only those relative at the biggest
correlation length of each simulated tempering run
(to avoid introducing correlated data in the fit)
and discarding the data at $\xi_G < 3$
obtained by using $S_g$
(this is justified by the slow approach to scaling expected
when using $S_g$ \cite{CPNlatt_2}).
Then for the ${\rm CP}^{20}$ model we found
\begin{equation}
\chi_t \xi_G^2\,=\,0.0076(3)\;,
\label{reschi}
\end{equation}
to be compared with the value $\chi_t \xi_G^2\,=\,0.00744$
coming from Eq.\ (\ref{chipred}).
However,
the result (\ref{reschi}) still disagrees with the L\"uscher large-$N$
prediction (\ref{chipredluscher}), which would require
$\xi_G^2/\xi_{\rm w}^2\simeq 2/3$, while we found
$\xi_G^2/\xi_{\rm w}^2\simeq 0.91$.

To extract the string tension we calculated  the Creutz ratios
$\eta(R)\equiv \chi(R,R)$.
The Wilson loops were measured using improved
estimators obtained by replacing each $\lambda_{n,\mu}$ with
its average $\lambda_{n,\mu}^{\rm imp}$ in the field of its
neighbors \cite{CPNlatt}.
In Fig.\ \ref{CR-plot} we plot the rescaled Creutz ratios
$\eta_r(R)$ defined in Eq.\ (\ref{Crratres}).
Data show a good agreement with the large-$N$ prediction (\ref{sigmapred}).
The data for the ${\rm CP}^{20}$ model  shown in Fig.\ \ref{CR-plot}
were taken at $\beta=0.65$
and $\beta=0.67$. At larger $\beta$ the signals were too noisy
for distances larger than $d\simeq 2\xi_G$, for $d < 2\xi_{\rm G}$
the results were consistent with those shown in Fig. \ref{CR-plot}.

We checked asymptotic scaling, according to the two loop formula
\begin{equation}
f(\beta) = (2\pi\beta)^{2/N}\exp (-2\pi\beta)\;,
\label{fbeta}
\end{equation}
by analyzing the
quantity $M_G/\Lambda_{\rm g}=[\xi_Gf(\beta)]^{-1}$.
We also analyzed the data by using the $\beta_E$ scheme,
in which a new coupling $\beta_E$ is extracted from the energy
and inserted in the two loop formula (\ref{fbeta}) \cite{Parisi,Parisi2}.
In Fig. \ref{asysc-plot} we plot the values of $M_G/\Lambda_g$
obtained with the two actions $S_g$ and $S_g^{\rm Sym}$
and by using the standard and the $\beta_E$ schemes.
To report all data in terms of $\Lambda_g$, we used
the ratios of the $\Lambda$ parameters given
in Ref. \cite{CPNlatt_2}.

At smaller $N$ the $\beta_E$ scheme showed a notable
improvement in testing asymptotic scaling, giving
also quite different values with respect to the standard scheme.
For example, for the ${\rm CP}^1$ (or $O(3)$ $\sigma$) model and
by using $S_g$,
the $\beta_E$ scheme gave a determination of $M_G/\Lambda_g$
in agreement with its analytical prediction
(within errors of about 3\%), while the standard scheme was out
by about 30\% at $\xi\simeq 30$ \cite{CPNlatt_2}.
At $N=21$ the discrepancy among the different determinations
is still present, although it is reduced. Instead at $N=41$
it has almost disappeared and the result is in good agreement
with the large-$N$ prediction (\ref{xiG/Lambda}).

In conclusion, the results of the Monte Carlo simulations
at $N=21$ and $N=41$ show a quantitative agreement
with the large-$N$
predictions for those quantities which are analytical functions
of $1/N$ around $N=\infty$ and which can be expanded in powers
of $1/N$, such as $\xi_{\rm G}$, $\chi_t$ and $\sigma$.
On the other hand, the approach to the large-$N$ asymptotic
regime of the quantities involving the mass gap appears very
slow and the ${\rm CP}^{40}$ should be still outside
the region where the complete mass spectrum predicted
by the $1/N$ expansion \cite{Haber} could be observed.

\acknowledgements
I would like to thank Massimo Campostrini and Paolo Rossi for
interesting and useful discussions.

% ========================= REFERENCES =========================

% ========================= FIGURE CAPTIONS =========================

\figure{The ratio $\xi_G/\xi_{\rm w}$
versus $\xi_G$. The dashed line shows the large-$N$ prediction
(\ref{ratioxi}). The dotted lines are the results of a fit.
\label{ratio_xi-plot}}

\figure{Topological susceptibility
versus $\xi_G$. The dashed lines show the large-$N$ prediction
(\ref{chipred}).
\label{chit-plot}}

\figure{The quantity $\eta_r(R)\xi_G^2$
as a function of the physical distance $R/\xi_G$.
The dashed lines show the value of the string tension
predicted by the large-$N$ expansion:
$\sigma \xi_G^2 = \pi/N$.\label{CR-plot}}

\figure{Asymptotic scaling test for $\xi_G$.
The dashed lines show the large-$N$ prediction (\ref{xiG/Lambda}).
\label{asysc-plot}}

% ========================= TABLES =========================

\mediumtext
\begin{table}
\caption{Summary of the simulation runs by using the simulated
tempering algorithm.
The runs are labeled by the letters a,b,c,d.
We report: the action $S$ used in the simulation;
the minimum and maximum value of $\beta$, ``range'';
the difference between two contiguous values of $\beta$,
$\Delta \beta$; the number of $\beta_i$, $N_\beta$;
the acceptance in the updating of $\beta$, $A_\beta$;
the total number of iterations, ``stat''.
}
\label{ST-table}
\begin{tabular}{rrlrrr@{}lrrr}
\multicolumn{1}{c}{}&
\multicolumn{1}{r}{$N$}&
\multicolumn{1}{c}{$S$}&
\multicolumn{1}{r}{$L$}&
\multicolumn{1}{c}{range}&
\multicolumn{2}{c}{$\Delta\beta$}&
\multicolumn{1}{r}{$N_\beta$}&
\multicolumn{1}{c}{$A_\beta$}&
\multicolumn{1}{r}{stat}
\\  \tableline
a & 21 & $S_g$ & 48 & 0.61--0.70 & 0&.003 & 31 & 68\% & 400k \\
b & 21 & $S_g$ & 60 & 0.60--0.72 & 0&.003 & 41 & 62\% & 700k \\
c & 21 & $S_g^{\rm Sym}$ & 48 & 0.51--0.63 & 0&.003 & 41 & 64\% & 700k \\
d & 41 & $S_g$ & 42 & 0.50--0.60 & 0&.0025 & 41 & 64\% & 500k \\
\end{tabular}
\end{table}

\begin{table}
\caption{Summary of the measurements.
An asterisk indicates runs with the Symanzik improved action.
We use the notation ``m,$\gamma$'' for a stochastic mixture
of microcanonical and over-heat bath updating with relative
weigth $\gamma$ (see Ref.\ \cite{CPNlatt}) and ``S.T.'' for
the simulated tempering algorithm.
The letters a,b,c,d near the values of $\beta$ indicate the simulated
tempering run where the measures were performed.
$t$ is the percentage of time
spent by the system  at a given  value of $\beta$.
%``n.e.'' labels the runs taken from Ref. \cite{CPNlatt},
%which did not sample correctly the topological sectors.
}
\label{datarun-table}
\begin{tabular}{rr@{}lrlr@{ }lr@{}lr@{}l}
\multicolumn{1}{r}{$N$}&
\multicolumn{2}{c}{$\beta$}&
\multicolumn{1}{r}{$L$}&
\multicolumn{1}{c}{Algor.}&
\multicolumn{2}{c}{stat}&
\multicolumn{2}{c}{$E$}&
\multicolumn{2}{c}{$\chi_{\rm m}$}
\\  \tableline
21 &  0&.65    & 42 & m,1 &  100k&  &  0&.7995(1)  &  12&.14(3) \\
%21 &  0&.70 n.e. & 48 & m,1 &   60k&  &  0&.7392(1)   &  19&.53(4) \\
%21 &  0&.75 n.e. & 60 & m,1 &   40k&  &  0&.6875(1)   &  31&.71(11) \\
21 &  0&.67 a  & 48 & S.T.& 400k& t=3.3\% &  0&.7741(1)  &  14&.70(3) \\
21 &  0&.70 a  & 48 & S.T.& 400k& t=3.3\% &  0&.7391(1)   &  19&.54(5)\\
21 &  0&.69 b  & 60 & S.T.& 700k& t=2.0\% &  0&.7504(1)  &  17&.73(3) \\
21 &  0&.72 b  & 60 & S.T.& 700k& t=1.7\% &  0&.7174(1)  &  23&.65(6) \\
\tableline
21 &  0&.60 $^*$ & 42 & m,1 & 150k&  & 0&.8582(1)  & 15&.08(3) \\
21 &  0&.60 $^*$ c&48& S.T. & 700k& t=2.4\% & 0&.8583(1)  & 15&.04(3)\\
21 &  0&.63 $^*$ c&48& S.T. & 700k& t=2.3\% & 0&.8161(1)  & 19&.89(4) \\
\tableline
41 &  0&.57  d &42& S.T. & 500k& t=2.4\% & 0&.8890(1)  & 7&.756(8)\\
41 &  0&.60  d &42& S.T. & 500k& t=1.8\% & 0&.8454(1)  & 10&.105(15)\\
\end{tabular}
\end{table}

\begin{table}
\caption{Results for the ${\rm CP}^{20}$ and the ${\rm CP}^{40}$
models.}
\label{some-table}
\begin{tabular}{rr@{}lrr@{}lr@{}lr@{}lr@{}l}
\multicolumn{1}{r}{$N$}&
\multicolumn{2}{c}{$\beta$}&
\multicolumn{1}{r}{$L$}&
\multicolumn{2}{c}{$\xi_G$}&
\multicolumn{2}{c}{$\xi_G/\xi_{\rm w}$}&
\multicolumn{2}{c}{$\chi_{\rm t}^{\rm g}\xi^2_G$}&
\multicolumn{2}{c}{$\beta^2Z_P$}
\\ \tableline
21 &  0&.65  & 42 &  2&.69(2)  & 0&.949(4)  & 0&.0088(5) &  0&.708(8) \\
%21 &  0&.70 n.e. & 48 & 3&.71(2) & 0&.950(4)  &  &---      &  0&.693(4) \\
%21 &  0&.75 n.e. & 60 & 5&.12(4) & 0&.947(7)  &  &---      &  0&.680(8) \\
21 &  0&.67 a & 48 & 3&.10(2)  & 0&.952(4)  & 0&.0074(6) &  0&.687(7) \\
21 &  0&.70 a & 48 & 3&.72(2)  & 0&.955(4)  & 0&.0081(8) &  0&.694(6) \\
21 &  0&.69 b & 60 & 3&.49(2)  & 0&.954(3)  & 0&.0083(7)&  0&.693(8) \\
21 &  0&.72 b & 60 & 4&.23(2)  & 0&.952(4)  & 0&.0078(7) &  0&.685(7) \\
\tableline
21 &  0&.60$^*$   & 42 & 2&.887(13) & 0&.957(5) & 0&.0070(8) & 0&.651(6) \\
21 &  0&.60$^*$ c & 48 & 2&.872(15) & 0&.961(3) & 0&.0075(4) & 0&.656(6) \\
21 &  0&.63$^*$ c & 48 & 3&.506(16) & 0&.955(3) & 0&.0075(6) & 0&.643(11) \\
\tableline
41 &  0&.57 d & 42 & 2&.011(10) & 0&.926(4) & 0&.0044(4) & 0&.623(7) \\
41 &  0&.60 d & 42 & 2&.431(11) & 0&.930(8) & 0&.0036(4) & 0&.616(7) \\
\end{tabular}
\end{table}

\end{document}